\documentclass[conference,10pt]{IEEEtran}
\usepackage[utf8]{inputenc}
\usepackage[T1]{fontenc}
\usepackage{mathtools}
\usepackage[thinc]{esdiff}
\usepackage{commath}
\usepackage{mathtools}
\usepackage{amsthm}
\usepackage{dsfont}
\allowdisplaybreaks
\usepackage{amsmath,amsfonts,amssymb}
\usepackage[ruled, linesnumbered]{algorithm2e}
\usepackage[noend]{algpseudocode}
\usepackage{float}
\usepackage{graphicx}
\usepackage{mathtools}
\usepackage[thinc]{esdiff}
\usepackage{epsf}
\usepackage{epsfig, amsmath, amssymb, graphicx}
\usepackage{subcaption}
\usepackage{multirow}
\usepackage{cite}
\usepackage[english]{babel}
\DeclareGraphicsExtensions{.pdf, .bmp, .eps, .ps, .jpeg, .png}

\long\def\comment#1{}

\usepackage{diagbox}

\usepackage{tabularx}
\newcolumntype{x}[1]{>{\centering\arraybackslash}p{#1}}
\newcolumntype{Y}{>{\small\centering\arraybackslash}X}

\title{Integrating Homomorphic Encryption and Trusted Execution Technology 
for Autonomous and Confidential Model Refining in Cloud}

\author{Pinglan Liu, Wensheng Zhang\\
Department of Computer Science, Iowa State University, 
Ames, Iowa, USA 50011\\
E-mail: \{pinglan,wzhang\}@iastate.edu}

\begin{document}
\maketitle


\begin{abstract}

With the popularity of cloud computing and machine learning,
it has been a trend to outsource machine learning processes (including 
model training and model-based inference) to cloud. 
By the outsourcing, 
other than utilizing the extensive and scalable resource offered by the cloud service provider,
it will also be attractive to users 
if the cloud servers can manage the machine learning processes
autonomously on behalf of the users. 
Such a feature will be especially salient when the machine learning is expected to 
be a long-term continuous process and the users
are not always available to participate.
Due to security and privacy concerns,
it is also desired that the autonomous learning preserves the confidentiality of users' data and models involved. 
Hence, in this paper, we aim to design a scheme that 
enables autonomous and confidential model refining in cloud. 
Homomorphic encryption and trusted execution environment technology can protect confidentiality for autonomous computation,
but each of them has their limitations respectively and they are complementary to each other. 
Therefore, we further propose to integrate these two techniques in the design of the model refining scheme. 
Through implementation and experiments,
we evaluate the feasibility of our proposed scheme.
The results indicate that, 
with our proposed scheme the cloud server can autonomously refine an encrypted model with newly provided encrypted training data
to continuously improve its accuracy. 
Though the efficiency is still significantly lower than the baseline scheme that
refines plaintext-model with plaintext-data, 
we expect that it can be improved by fully utilizing the higher level of parallelism and the computational power of GPU at the cloud server. 

\end{abstract}



\section{Introduction}
\label{sec:intro}


Since a decade ago, the inception and popularity of cloud computing paradigm
has changed the way people manage their data, processes, and IT infrastructure.
Many organizations and individuals have utilized
the cloud-based, instead of self-managed on-premise,
hardware/software infrastructures for data storage and processing.
The cloud users can enjoy convenient access to scalable resources while relieved from the burden of
managing their IT resource with the expectations of efficiency, reliability and security.
Such advantages are especially attractive to 
the users such as small/middle-size businesses and individuals 
who lack the required expertise and/or cannot afford 
the costs for managing the data, processes, and infrastructure on-premise.
Meanwhile, machine learning has been revolutionizing
the processing of image, audio and natural language and many other application domains.
People enjoy unprecedential benefits brought by 
the results of ubiquitous machine learning from extensive kinds of data.

Under the influence of the above two sweeps,
it has been a natural trend to integrate cloud computing with machine learning.
That is, the machine learning processes (including
{\em training} models from data and using models to {\em infer} from data)
can be outsourced to the computational platform in cloud, and
the data used by the training and inference processes
can be outsourced to the storage platform in cloud.
By the outsourcing, 
other than utilizing the extensive and scalable resource offered by the cloud server,
it will also be attractive to users 
if the cloud servers can manage the machine learning processes {\em autonomously} 
on behalf of the users, with perhaps only minimal intervention from them.
Such a feature will be especially salient when 
the machine learning is expected to be a {\em long-term} and {\em continuous} process, where
users keep adding new data every now and then, and
the cloud server should use the newly-added data to continuously refine the existing model in the {\em background},
such that the model keeps evolving over the time.
With this feature,
users can also enjoy the flexibility and convenience of
no-need to participate in the continuous learning; in particular,
they do not have to remain online all the time.
In this research,
we are interested in attaining the above feature of
cloud-based autonomous and continuous model refining.

The integration of cloud computing and machine learning
does not come without challenges.
Confidentiality protection is one of them.
People who provide data for training or inference often
want to keep their data confidential or 
have their own privacy protected. 
(Note that, confidentiality usually implies privacy protection.)
People who pay for model training usually want to keep their model confidential.
The cloud-based infrastructure, however, is not fully trusted to protect the confidentiality of data or model.
Particularly, cloud servers may be compromised by outside attackers, 
as evident by the increasing media coverage of data breach accidents.
Also, the regulations and procedures inside the cloud servers are often not transparent enough,
thus confidentiality may also be compromised due to intentional or accidental misbehavior
of their employees or systems.
Mechanisms should be in place to protect the confidentiality.

Over the past years,
extensive research has been conducted with the aim to protect data and/or model confidentiality,
mostly in the context of model-based inference.
Roughly, the approaches used by these efforts include 
multi-party computation (MPC), 
trusted execution environment technology (TEE), and
homomorphic encryption (HE),
and each of these approaches has its pros and cons. 

Among these approaches, 
the MPC-based solutions~\cite{mohassel2017secureml, rouhani2018deepsecure, liu2017oblivious, juvekar2018gazelle, mishra2020delphi, chandran2019ezpc, riazi2019xonn, riazi2018chameleon,mohassel2018aby3, wagh2020falcon,baryalai2016towards}
have been studied the most extensively, 
due to its computational efficiency. 
However, such approaches require multiple parties to interact with each other when
the outsourced computation is being ongoing; thus, 
higher communication overhead and network latency can be incurred. 
In addition, some proposed solutions~\cite{patra2020blaze,koti2021swift,chaudhari2019trident,byali2019flash}
expect the involved parties not to collude. 

With the support from hardware technologies such as Intel SGX, 
the TEE-based solutions~\cite{zhang2021citadel,natarajan2021chex} 
establish secure execution environment (called enclave for Intel SGX) to 
run outsourced confidential computation.
If the secure environment has large enough memory space to
execute the confidential computation without incurring frequent page swapping, 
this approach can attain good efficiency in terms of both computation and communication, 
because the computation can be performed directly over plaintexts 
and there is no need for interactions between multiple parties.
However, this approach cannot fully utilize the resource available at cloud server.
Specifically, the memory and processing resources available for secure execution environment 
is usually a small portion of all the resource available at a cloud server; for example,
though some new-generation Intel CPUs~\cite{intelXeonSGX} for servers support SGX memory in the range of 8GB-512GB,
it is still significantly smaller than the size of the regular memory (up to 6TB) that can be supported. 
Also, the secure execution environment cannot utilize the computationally-powerful GPUs yet. 
Besides, this approach can suffer from side-channel attacks~\cite{wang2017leaky, van2018foreshadow, chen2019sgxpectre}.

The HE-based approach utilizes the feature of Homomorphic encryption that,
computation can be conducted over encrypted data
and thus the confidentiality of data is preserved without sacrificing its utility. 
Like the TEE-based approach, 
the HE-based approach does not require interactions between multiple parties during computation
and thus is more communication efficient.
Moreover, 
the HE-based approach does not suffer from side-channel attacks. 
However, homomorphic encryption/decryption and 
computation over homomorphically-encryption data are still computational expensive,
though much advancement has been made to improve the encryption algorithms~\cite{brakerski2014efficient, simd, cheon2017homomorphic}
and to utilize GPU for accelerating the computation~\cite{al2020multi, lee2022privacy, turan2020heaws}.
In particular, 
as more and more computations are conducted over homomorphically-encrypted data,
noises accumulate at the resulting ciphertext. 
Note that, 
the problem of noise accumulation is more severe in confidential model training/refining than 
in confidential model-based inference. 
With the inference process,
encrypted inputs are fed into and then propagate forward through an encrypted model
without updating the parameters of the model; hence, noises only accumulate
during one pass of the propagation. 
However, with the training/refining process,
encrypted training data propagate forward and then backward through
the encrypted model, where the parameters of the model are updated 
during the backward propagation; hence, the noises introduced to the parameters at one propagation
carry on to the subsequent propagations. 
Therefore, for HE-based confidential training/refining, 
the noises should be removed periodically to allow more computation to be conducted. 
If bootstrapping is used frequently for this purpose, even higher cost can be incurred.

From the above survey, we have the following observations. 
First, due to its requirement for frequent interactions among multiple parties during computation,
the MPC-based approach is not effective to 
support autonomous and continuous refining of a model, 
though it could be used in intensive training occasions where 
the parties (e.g., the users) are assured to participate. 
Second, the TEE-based and the HE-based approaches 
are complementary to each other, and 
can be integrated to compose an effective solution that
protects confidentiality in autonomous and continuous model refining. 
Based on these observations, 
we propose a scheme that integrates homomorphic encryption and trusted execution technology
for autonomous and confidential model refining, 
which is summarized as follows. 

We consider a cloud server which includes two parts,
the trusted execution environment (TEE) and 
the regular execution environment (REE). 
The TEE is trustworthy and can be attested by any other party interacting with it.
It provides the basic function of initializing the system, 
which particularly includes generating and distributing the keys for 
homomorphic encryption and 
for computation over homomorphically-encrypted data.
In addition, the TEE also provides the service for re-encrypting data,
where already-encrypted data are decrypted and then encrypted again so that 
the noises accumulated in the encrypted data can be reduced.
Such a design of the TEE is for the purpose of
minimizing its functionality, 
and hence minimizing the need for resource and 
the chance of being attacked via side channels.

A client of the above cloud server is 
one user or one group of users,
who already has a base model but 
needs to keep refining the model for long term. 
Here, the base model may have been trained 
either directly at the client or 
via multi-party computation. 
As an extreme case, 
the base model could be just be an initial model that
has not been trained with any data yet.
The base mode should be homomorphically-encrypted 
with a key provided by the server's TEE and 
then outsourced to the server.
After outsourcing the base model, 
the users submit their new data, 
which should also be homomorphically-encrypted,
to the server every now and then.
Upon receiving a batch of new data,
the server should conduct additional training to the existing model for the purpose of model refining. 
The algorithm for training has been carefully designed such that,
it can conduct each round of training efficiently by 
minimizing the times of computationally-heavy operations,
and it can also make use of TEE's re-encryption service 
to efficiently reduce noises in the ciphertext periodically. 

We have implemented the system, and 
evaluated the feasibility of the design and 
the performance via experiments over a moderate computation platform. 
The results show that, with the proposed scheme,
the cloud server can work autonomously to 
refine an encrypted base model with a new batch of encrypted data provided by clients,
without intervention from the clients. 
The refining process can gradually increase the accuracy of the model,
at a rate lower than but comparable to the baseline scheme where 
the same base model (but in plaintext) is refined based on the same batch of data (in plaintext)
using the default Pytorch code. 
As the computation is all over ciphertexts,
higher computational costs are incurred. 
Note that, the experinment only uses a single CPU core and does not utilize GPU.

In the rest of the paper,
we present the problem description and background in Section II,
and our proposed solution in Section III.
Section IV reports the evaluation results,
and Section V briefly surveys the related works.
Finally, Section VI concludes the paper.

\begin{figure*}[htb]
    \centering
    \caption{Framework of Proposed Scheme. 
    }
    \label{fig:framework}
    \includegraphics[width=0.8\linewidth]{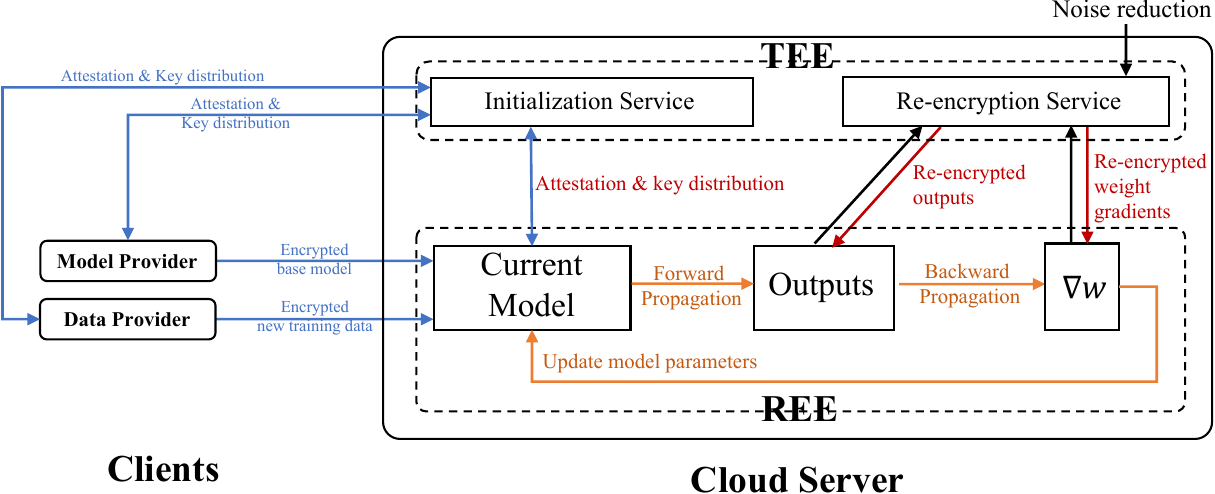}
\end{figure*}

\section{Problem Description and Background}
\label{sec:prelim}

In this section, we present 
the system model, 
the problem definition,
the CNN model that is used in our scheme,
the leveled homomorphic encryption (LHE) primitives utlized by our scheme, 
and the LHE-based confidential inference scheme~\cite{liu2022towards} that 
our proposed refining scheme is based on. 

\subsection{System Model and Problem Description}
\label{subsec:syst-model}

The system we propose consists of a cloud server and one or multiple clients. 
The server contains a trusted execution environment (TEE), 
such as an Intel SGX enclave, 
and an untrusted regular execution environment (REE) 
that is rich in resources and includes all resource outside of the TEE.
The TEE is trustworthy and can be attested by any other party that interests with it. 

We consider two types of clients, 
model provider and data provider.
Note that these two types are conceptual; 
in reality, a physical client can be both a model provider and a data provider at the same time.  

The model provider already owns a 
model called the base model,
which can have been trained based on a dataset locally by the client or together with other parties.
As an extreme case, 
the model can also be just an initial model that has not been trained yet.

A data provider owns a small dataset at a time.
This dataset may have a different distribution from the dataset that had been used to train the base model.  
The data provider 
submits its dataset to the cloud server, 
and wants the server to 
refine (i.e., train) 
the existing model to fit with the new dataset. 
For data and model confidentiality, 
it is critical that the data is only disclosed in plaintext to its providing client, 
and the trained model parameters are only revealed in plaintext to the model provider. 
Nevertheless, we permit the disclosure of hyper-parameters of the model, 
such as the number of layers and nodes on each layer.

\subsection{CNN Model}
\label{subsec:cnn-model}

In this work,
we use the CNN model as the example of model to refine. 
%
%
Specifically, we consider the CNN model with $c$ convolutional layers (CLs)
and $f$ fully-connected layers (FLs). 

For each CL $l\in\{0,\cdots,c-1\}$,
the channel number is denoted as $\alpha_l$.
The layer also has an input matrix for each channel, 
consisting of $\beta_l\times\beta_l$ elements, 
where $\beta_l$ is defined as the side of the input matrix. 
Additionally, each channel has $\epsilon_l$ filters, 
where the side of each filter is $\gamma_l$ and the stride is $\delta_l$.
The filter and gradient of a filter of layer $l$ 
is denoted as $\hat{F}^{l,k}_{i,x,y}$ and
$\Tilde{F}^{l,k}_{i,x,y}$ respectively
where $k\in\{0,\cdots,\epsilon_l-1\}$,
$i\in\{0,\cdots,\alpha_l-1\}$ and
$x,y\in\{0,\cdots,\gamma_l\}$.

Each fully-connected layer $l\in\{0,\cdots,f-1\}$ 
is characterized by the number of input and output neurons, 
denoted as $\iota_l$ and $o_l$, respectively. 
As a result, the weight matrix for this layer, 
denoted as $M^{(l)}$, has dimensions $\iota_l\times o_l$.
The weight gradient for this layer is denoted as $\Tilde{M}^{(l)}$.
The output for each layer either in CL or FL $l$ is denoted as $\hat{C}^{(l+1)}$ 
while the output gradient in backpropagation is denoted as $\Tilde{C}^{(l+1)}$.

\subsection{Leveled Homomorphic Encryption (LHE) Primitives}

We adopt an asymmetric leveled homomorphic encryption (LHE) scheme 
that allows computation on encrypted data (ciphertexts) 
up to a certain predefined {\em level}. 
Our scheme also employs ciphertext packing, 
enabling multiple values to be encoded 
and encrypted into a single ciphertext 
and operations to be performed in a SIMD manner. 
The LHE scheme is composed of the following primitives,
where we follow the notations used in~\cite{liu2022towards}: 

\begin{itemize}
    \item 
    $(sk,pk,evk,{\cal S})\leftarrow KeyGen(1^\lambda, {\cal L})$: 
    with security parameter $\lambda$ and the highest level of encryption ${\cal L}$ as inputs, 
    the primitive for key generation outputs 
    a secret key $sk$, 
    a public key $pk$, 
    an evaluation key $evk$ and
    the the slot number ${\cal S}$ for each ciphertext.
    Particularly, ${\cal S}$ indicates the maximum number of scalar values that 
    can be encoded and encrypted within a single ciphertext.

    \item
    $ct\leftarrow Enc_{pk}(\vec{pt})$: 
    provided public key $pk$ and 
    a plaintext vector $\vec{pt}=(pt_0,\cdots,pt_{{\cal S}-1})$
    of ${\cal S}$ elements, 
    the primitive for encryption outputs a ciphertext $ct$.
    Note that a newly encrypted ciphertext has the smallest noise or the highest level (i.e., ${\cal L}-1$);
    we denote this as $ct.level = {\cal L}-1$.

    \item
    $\vec{pt}\leftarrow Dec_{sk}(ct)$: 
    provided secret key $sk$ and a ciphertext $ct$, 
    the primitive for decryption outputs plaintext vector $\vec{pt}$ s.t. $Enc_{pk}(\vec{pt})=ct$.

    \item 
    $ct'\leftarrow ct_1\oplus ct_2$: 
    provided ciphertexts $ct_1$ and $ct_2$, 
    the primitive for addition outputs ciphertext $ct'$ s.t. $Dec_{sk}(ct')=Dec_{sk}(ct_1)+Dec_{sk}(ct_2)$. 
    Here, the $+$ operator stands for element-wise 
    addition between two vectors; that is, 
    if $Dec_{sk}(ct_1)=\vec{pt}_1 = (pt_{1,0},\cdots, pt_{1,{\cal S}-1})$ and 
    $Dec_{sk}(ct_2)=\vec{pt}_2 = (pt_{2,0},\cdots, pt_{2,{\cal S}-1})$,
    then $Dec_{sk}(ct_1)+Dec_{sk}(ct_2) = (pt_{1,0}+pt_{2,0},\cdots, pt_{1,{\cal S}-1}+pt_{2,{\cal S}-1})$.
    Note that, applying this primitive aggregate the noises in the operands $ct_1$ and $ct_2$ to $ct'$, 
    and $ct'$ has the level which is the smaller one between those of $ct_1$ and $ct_2$;
    specifically, $ct'.level = \min\{ct_1.level, ct_2.level\}$. 

    \item 
    $ct'\leftarrow ct_1\otimes ct_2$: 
    provided ciphertexts $ct_1$ and $ct_2$, 
    the primitive for multiplication between ciphertexts 
    outputs ciphertext $ct'$ s.t. $Dec_{sk}(ct')=Dec_{sk}(ct_1)\times Dec_{sk}(ct_2)$. 
    Here, the $\times$ operator stands for element-wise 
    multiplication between two vectors; that is, 
    if $Dec_{sk}(ct_1)=\vec{pt}_1 = (pt_{1,0},\cdots, pt_{1,{\cal S}-1})$ and 
    $Dec_{sk}(ct_2)=\vec{pt}_2 = (pt_{2,0},\cdots, pt_{2,{\cal S}-1})$, then 
    $Dec_{sk}(ct_1)\times Dec_{sk}(ct_2) = (pt_{1,0}\times pt_{2,0},\cdots, pt_{1,{\cal S}-1}\times pt_{2,{\cal S}-1})$.
    Note that, applying this primitive results in larger noise at the resulting ciphertext than that of every operands
    and hence the level of the result is lower than that of every operands;
    specifically, $ct'.level = \min\{ct_1.level, ct_2.level\}-1$.
    
    \item 
    $ct'\leftarrow CMult(ct,\vec{pt})$: 
    provided ciphertext $ct$ and plaintext $\vec{pt}$, 
    the primitive for multiplication between plaintext and ciphertext 
    outputs ciphertext $ct'$ s.t. $Dec_{sk}(ct')=\vec{pt}\times Dec_{sk}(ct)$.
    Note that, applying this primitive also results in larger noise at the resulting ciphertext than that of the ciphertext operand
    and hence the level of the result is lower than that of the operand;
    specifically, $ct'.level = ct.level -1$.
    
    \item 
    $ct'\leftarrow Rot(ct,m)$: 
    provided ciphertext $ct$ that encrypts $\vec{pt}=(pt_0,\cdots,pt_{{\cal S}-1})$ and integer $m<{\cal S}$, 
    the primitive for rotation outputs $ct'$ which is ciphertext for $(pt_{m},\cdots,pt_{{\cal S}-1},pt_0,\cdots,pt_{m-1})$.
    Note that, applying this primitive does not change the level; that is, $ct'.level = ct.level$.

\end{itemize}
Our design and implementation utilize CKKS~\cite{cheon2017homomorphic},
which offers all of the aforementioned primitives required by our proposed system.
In addition, our system also uses the following primitive for re-encryption,
which can reduce the noises of a ciphertext and recover its level to the highest level.
\begin{itemize}
    \item 
    $ct'\leftarrow Reencrypt_{sk,pk}(ct)$:
    provided secret key $sk$, publick key $pk$ and ciphertext $ct$ at certain level between 0 and ${\cal L}-1$,
    the primitive for re-encrypton outputs $ct'$ s.t., $ct'=Enc_{pk}(Dec_{sk}(ct))$.
    Here, $ct'.level = {\cal L}-1$ regardless of the level of $ct$.
\end{itemize}

\subsection{LHE-based Inference Scheme}

A round of learning process includes a forward propagation and a backward propagation,
where the forward propagation is the same as the model-based inference process.
Our proposed scheme reuses the LHE-based inference scheme 
proposed by Liu and Zhang~\cite{liu2022towards}
for the forward propagation, which is reviewed as follows. 


We first review 
the framework proposed in \cite{liu2022towards},
which involves the TEE of the cloud server, 
the REE of the cloud server, 
a model provider, and 
a data provider.
Following the framework,
a process starts with the model provider attesting the TEE
as well as sharing its secret key and the hyper-parameters of the CNN model 
with the TEE upon a successful attestation. 
Based on the CNN model's architecture, 
the TEE generates keys for asymmetric leveled homomorphic encryption 
via the GenKey algorithm and securely sends the public key to the model provider. 
The model provider encrypts its model parameters and 
uploads the encrypted parameters to the REE. 
Then, the TEE sends the public key and the evaluation key to the REE.
The data provider also attests the TEE 
and shares its secret key with the TEE after a successful attestation. 
The TEE securely sends the public key to the data provider, 
who then encodes and encrypts its data before uploading the encrypted data to the REE. 
The REE conducts the inference based on the encrypted model and encrypted data, 
and sends the encrypted inference result to the TEE for decryption. 
Finally, the TEE decrypts the result and returns it to the data provider 
after further encrypting it with the data provider's secret key. 
The data provider decrypts the inference result using its secret key.

Next, we review the computation conducted by REE, 
which is the LHE-based inference process and 
the same as the forward propagation of a learning process.
The proposed LHE-based inference scheme consists of the methods for encoding data and filters,
as well as for performing forward propagation over convolutional layers and fully-connected layers.
To enable efficient propagation through all $c$ convolutional layers
while only requiring packing and encryption at the beginning,
each input matrix should be encoded and encrypted 
as if it were part of a single virtual layer that encompasses all $c$ layers in order. 
This layer is referred to as the "combined layer", 
and the parameters $\tilde{\delta}_0$ and $\tilde{\gamma}_0$ are used
to denote its combined stride and kernel side, respectively 
(where $\tilde{\gamma}_0^2$ represents the number of encrypted inputs for each channel).
To ensure consistency with different types of inputs 
and accommodate $f$ fully-connected layers, 
two weight matrix encoding methods are introduced. 
In the following part, 
we provide a detailed examination of these methods.

\subsubsection{Encoding and Encrypting Inputs}

To take full advantage of the $\cal {S}$ slots, 
$n$ inputs are encoded simultaneously, 
with each input image having a dimension of $\alpha_0\times\beta_0\times\beta_0$. 
These values are then encrypted into $\alpha_0\times\tilde{\gamma_0}^2$ ciphertexts, 
indexed by $(i,u,v)$ for $i\in\{0,\cdots,\alpha_0-1\}$ 
and $(u,v)\in\{0,\cdots,\tilde{\gamma}_0-1\}^2$. 
Specifically, $\hat{C}^{(0)}_{i,u,v}$ encrypts $\vec{I}_{i,u+s\times\tilde{\delta}_0,v+t\times\tilde{\delta}_0}$, where $(s,t)\in\{0,1,\cdots\tilde{\beta}_0-1\}$ 
and $\vec{I}_{i,u,v} = ({I}_{i,u,v}^{(0)},~\cdots,~{I}_{i,u,v}^{(n-1)}).$

\subsubsection{Encoding and Encrypting Filters}

To accommodate the encoding and encryption of input data, 
all values within the same ciphertext must be multiplied by the same filter element. 
Thus, each element $F_{i,j}$ of a filter $F$ must be replicated $n\times\tilde{\beta}_0^2$ times 
and then encoded and encrypted into a ciphertext $\hat{F}_{i,j}$.

\subsubsection{Forward Propagation through CL $l$}

Each propagation layer consists of multiple input ciphertexts,
which are of dimension $\tilde{\gamma}_l\times\tilde{\gamma}_l$ for each channel. 
These input ciphertexts are denoted as $\hat{C}^{(l)}_{i,u,v}$ 
for every channel $i$ and every pair $(u,v)$. 
Similarly, the encrypted elements of the filters are denoted as $\hat{F}^{(l,k)}_{i,x,y}$ 
for every channel $i$, filter index $k$, and element index $(x,y)$ of the filter matrix. 
The convolutional operations are performed between these input ciphertexts and encrypted filter elements. 
The neurons undergo a square activation function, and the resulting outputs become the input of the next layer. 
In the final convolutional layer, the output values of the same channel are packed together into one single ciphertext.

\subsubsection{Forward Propagation through FL $l$}

As explained earlier, the encoding technique employed by the model 
orders input values with the same offset but originating from $n$ distinct inputs consecutively. 
This sequence of $n$ values is recognized as a {\em parallel input set (pi-set)}. 
Consider a certain FL layer $l$ where $\iota'_l$ input ciphertexts exist, 
and each ciphertext encrypts $\iota''_l$ pi-sets. 
Then, the total number of pi-sets present in layer $l$ is $\iota_l=\iota'_l\times\iota''_l$.
Two packing methods are proposed based on the input types. 
In type I packing, 
each input ciphertext encrypts multiple pi-sets without replication, 
while in type II packing, 
each input ciphertext encrypts only one pi-set but is replicated multiple times.
To be more specific, 
assume the weight matrix $M^{(l)}$ is of dimension $\iota_l\times o_l$.
Type I packing method encodes the weights associated with each output neuron into the same ciphertext. 
During forward propagation, 
this method generates $o_l$ ciphertexts as outputs, 
where each ciphertext contains a pi-set that appears multiple times. 
These $o_l$ ciphertexts serve as inputs for the Type II encoding method. 
In Type II encoding, 
weights connected to the same input neuron are encrypted into the same ciphertext, 
resulting in output ciphertexts in the same format as Type I. 
Therefore, Type I and Type II encoding alternate in the forward propagation process.



\section{Proposed Scheme}
\label{sec:solution}

In this section,
we first present an overview of our proposed scheme,
which is followed by detailed description of the designs for backward propagation 
through fully-connected layers and convolutional layers. 

\subsection{Overview}
\label{sec:overview}

Figure~\ref{fig:framework} 
provides an overview of our proposed scheme,
which is explained as follows. 
The 
proposed scheme involves four parties: 
the TEE of the cloud server, 
the REE of the cloud server,
a model provider, and a data provider. 

The TEE is first deployed with appropriate security parameters,
based on which it initializes itself by generating
the keys for homomorphic encryption/decryption and evaluation.
Then, it starts two services,
{\em initialization service} and {\em re-encryption service}
for other parties.

When the autonomous refining service is deployed to the REE,
the service interacts with the TEE to conduct the attestation;
once the attestation succeeds, it should receive from the TEE 
the keys needed for conducting computation over homomorphically-encrypted data.
After then, the service at REE waits for the joining of the model and data providers.

When a model provider joins,
it should interact with the TEE for attestation; 
once the attestation succeeds, 
it should get the public key and use it to homomorphically-encrypt the parameters of its model.
Then the encrypted model should be provided to the refining server at REE as the base model.

Every time when a data provider joins with a new dataset to contribute, 
it works as follows.
If it is the first time for it to join the system,
it should interact with the TEE to attest it;
once the attestation succeeds, it should get the public key for homomorphic encryption. 
With the public key available, which is obtained either this time or earlier,
the data provider should properly pack and encode its data,
according to the packing algorithm in \cite{liu2022towards} which is also reviewed in Section II. 
Then, it homomorphically-encrypts the packed data and sends the result to 
the refining service at REE to conduct a new pass of refining. 

A pass of refining includes multiple rounds of learning,
and each round includes a forward propagation and a backward propagation.
The process for forward propagation is similar to the process of inference based on an existing model,
and so we reuse the confidential inference algorithm in \cite{liu2022towards} for the purpose.
In the rest of the section,
we elaborate the confidential backward propagation,
which further includes the backward propagation through each fully-connected layer (FL) 
and each convolutional layer (CL).




\subsection{Backpropagation through Fully-connected Layer (FL) $l$}

Recall that there are two types of inputs and encoding methods for the FL layer.
Here, we highlight our proposed backpropagation algorithms for outputs and weight gradients.
Note that, each FL is followed by a square activation function.
The backpropagation for outputs computes the gradients of the activation function first.

\subsubsection{Backpropagation with Type I Input}

\begin{algorithm}[htb]
\SetAlgoLined
\caption{Backward Propagation through Fully-connected Layer $l$ (with Type I Input)}
\label{alg:bwd-full-layer-I}
\scriptsize
\For{$j\in\{0,\cdots,o_l-1\}$}{
    $\tilde{C}^{(l+1)}_j \leftarrow 2\otimes\tilde{C}^{(l+1)}_{j}$
    \Comment{activation derivation}
}
\For{$i\in\{0,\cdots,\iota'_l-1\}$}{
    $\tilde{C}^{(l)}_i\leftarrow 0$; \Comment{each input element $i$}\;
    \For{$j\in\{0,\cdots,o_l-1\}$}{
        $\tilde{C}^{(l)}_i \oplus= \tilde{C}^{(l+1)}_{j}\otimes\hat{M}^{(l)}_{j,i}$
    }
}
\end{algorithm}

In type I input, multiple pi-sets without duplicates are contained in a single ciphertext, 
and the weights connected to the same output neuron are encrypted into that same ciphertext. 
Each output neuron itself is a ciphertext containing one pi-set which duplicates for multiple times.
For FL layer $l$ with $\iota'_l$ input ciphertexts, 
denoted as $\hat {C}^{(l)}_i$ for $i\in\{0,\cdots,\iota'_l-1\}$,  
the output gradient for layer $l$ is denoted as $\tilde {C}^{(l)}_i$.
The weight matrix is represented as $\hat{M}_{j, i}^{(l)}$ for $j\in\{0,\cdots,o_l-1\}$, 
where $o_l$ refers to the number of output neurons. 
Then, the output gradients $\tilde {C}^{(l)}_i$ are calculated 
as the LHE-based multiplication of 
the output gradients for layer $l+1$ and the encrypted weight matrix. 
Finally, the multiplications are aggregated for the same input ciphertext.
The algorithmic process is formally present in Algorithm~\ref{alg:bwd-full-layer-I}.

\subsubsection{Backpropagation with Type II Input}

\begin{algorithm}[htb]
\SetAlgoLined
\caption{Backward Propagation through Fully-connected Layer $l$ (with Type II Input)}
\label{alg:bwd-full-layer-II}
\scriptsize
\For{$j\in\{0,\cdots,\lceil\frac{o_l\cdot n}{\cal S}\rceil-1\}$}{
    $\tilde{C}^{(l+1)}_j \leftarrow 2\otimes\tilde{C}^{(l+1)}_{j}$
    \Comment{activation derivation}
}
\For{$i\in\{0,\cdots,\iota_l-1\}$}{
    $\tilde{C}^{l}_i\leftarrow 0$; \Comment{each input element $i$}\;
    \For{$j\in\{0,\cdots,\lceil\frac{o_l\cdot n}{\cal S}\rceil-1\}$}{
        $\tilde{C}^{(l)}_i \oplus= \tilde{C}^{(l+1)}_{j}\otimes\hat{M}^{(l)}_{j,i}$
    }
    \For{$j\in\{1,\cdots,\log(\frac{\cal S}{n})\}$}{
        $\tilde{C}^{(l)}_i \oplus= Rot(\tilde{C}^{(l)}_i,j\cdot n)$;
    }
}
\end{algorithm}

Type II inputs consist of input neurons each represented as a ciphertext, 
and each input ciphertext contains one pi-set that is duplicated for multiple times. 
The weights connected to the same input neuron are encoded jointly, 
resulting in the output ciphertext encrypting multiple pi-sets without duplication. 
Specifically, in layer $l$, 
the input ciphertext is denoted as $\hat{C}_i^{(l)}$ for $i\in\{0,\cdots,\iota_l-1\}$, 
and the weight is represented as $\hat{M}^{(l)}_{j,i}$ for $j\in\{0,\cdots,\lceil\frac{o_l\cdot n}{\cal S}\rceil-1\}$, 
where $\lceil\frac{o_l\cdot n}{\cal S}\rceil$ is the number of output ciphertexts. 
We can obtain the output gradients $\tilde{C}_i^{(l)}$ for layer $l$ 
by leveraging LHE-based 
multiplication 
of the output gradients in layer $l+1$ (i.e., $\tilde{C}_i^{(l+1)}$) and 
the encrypted weight $\hat{M}^{(l)}_{j,i}$. 
The algorithmic process is formally presented in Algorithm~\ref{alg:bwd-full-layer-II}.

\begin{algorithm}[htb]
\SetAlgoLined
\caption{Calculating Rotation Steps}
\label{alg:bwd-cpt-rotation}
\scriptsize
$\vec{R}\leftarrow 1$;
\Comment{initialize output}\;
\For{$i\in\{\lceil\log n\rceil,\cdots,0\}$}{
    \If{$p\geq 2^i$}{
        $p -= 2^i$;
        $\vec{R}_i = -1$\;
    }
}
\end{algorithm}

\begin{algorithm}[htb]
\SetAlgoLined
\caption{Updating Weight Gradients through Fully-connected Layer (FL) $l$}
\label{alg:bwd-weights-full-layer}
\scriptsize
\For{$j\in\{0,\cdots,o'_l-1\}$}{ \Comment{$o'_l=o_l$ for type I and $o'_l=\lceil\frac{o_l\cdot n}{\cal S}\rceil-1$ for type II}\;
    \For{$i\in\{0,\cdots,\iota'_l-1\}$}{
        $\tilde{M}^{(l)}_{j,i} = \tilde{C}^{(l+1)}_{j}\otimes\hat{C}^{(l)}_{i}$;
    }
}
\For{$j\in\{0,\cdots,o'_l-1\}$}{ 
    \For{$i\in\{0,\cdots,\iota'_l-1\}$}{
        \Comment{sum up gradients from $n$ parallel images to position $p$ for $\tilde{M}^{(l)}_{j,i}$}\;
        $p \leftarrow (j\cdot\iota'_l+i)\%n$\;
        $\vec{R} \leftarrow computeRotations(p, n)$; 
        \Comment{compute the rotate directions and steps to sum up gradients}\;
        \For{$k\in\{0,\cdots,\lceil\log n\rceil\}$}{
            $\tilde{M}^{(l)}_{j,i} \oplus= Rot(\tilde{M}^{(l)}_{j,i}, 2^k\cdot\vec{R}_k)$;
        }
    }
}
\end{algorithm}

\begin{algorithm}[htb]
\SetAlgoLined
\caption{Removing Accumulated Noise of Weights in Fully-connected Layer (FL) $l$}
\label{alg:bwd-weights-TEE-full-layer}
\scriptsize
\For{$j\in\{0,\cdots,o'_l-1\}$}{ 
    \For{$i\in\{0,\cdots,\iota'_l-1\}$}{
        \Comment{pack weight gradients $\tilde{M}^{(l)}_{j,i}$ together}\;
        $(p,~k)\leftarrow ((j\cdot\iota'_l+i)\%n,~(j\cdot\iota'_l+i)/n)$\;
        $\vec{U} \leftarrow [0,\cdots,\beta,\cdots,0,\cdots,\beta,\cdots]$;
        \Comment{for position $\{p,p+n,p+2n,\cdots,p+\lfloor\frac{{\cal S} -p}{n}\rfloor\cdot n\}$, $\vec{U}$ has value $\beta=\frac{lr}{n}$ where $lr$ is learning rate, otherwise 0}\;
        \If{not visited $k$}{
            $\tilde{PM}^{(l)}_k\leftarrow 0$;
            \Comment{initialize each output}\;
        }
        $\tilde{PM}^{(l)}_k \oplus= \tilde{M}^{(l)}_{j,i}\otimes\vec{U}$; 
    }
}
\For{$k\in\{0,\cdots,\lceil\frac{o'_l\cdot\iota'_l}{n}\rceil-1\}$}{
    \Comment{decrypt, decode, re-encode and re-encrypt the packed weights in TEE}\;
    $\vec{pt}\leftarrow Dec_{sk}(\tilde{PM}^{(l)}_k)$;
    $\tilde{PM}^{(l)}_k \leftarrow Enc_{pk}(\vec{pt})$;
}
\For{$j\in\{0,\cdots,o'_l-1\}$}{ 
    \For{$i\in\{0,\cdots,\iota'_l-1\}$}{
        \Comment{unpack weights $\tilde{PM}^{(l)}_k$ and recover $\tilde{M}^{(l)}_{j,i}$}\;
        $(p,~k)\leftarrow ((j\cdot\iota'_l+i)\%n,~(j\cdot\iota'_l+i)/n)$\;
        $\vec{U} \leftarrow [0,\cdots,1,\cdots,0,\cdots,1,\cdots]$;
        \Comment{for position $\{p,p+n,p+2n,\cdots,p+\lfloor\frac{{\cal S} -p}{n}\rfloor\cdot n\}$, $\vec{U}$ has value 1, otherwise 0}\;
        $\hat{M}^{(l)}_{j,i} \leftarrow \tilde{PM}^{(l)}_k\otimes\vec{U}$\; 
        $\vec{R} \leftarrow computeRotations(p, n)$; 
        \Comment{compute the rotation directions and steps to spread gradients}\;
        \For{$t\in\{0,\cdots,\lceil\log n\rceil\}$}{
            $\tilde{M}^{(l)}_{j,i} \oplus= Rot(\tilde{M}^{(l)}_{j,i}, -2^t\cdot\vec{R}_t)$;
        }
        $\hat{M}^{(l)}_{j,i} \oplus= \tilde{M}^{(l)}_{j,i}$;
        \Comment{update weight with gradients}
    }
}
\end{algorithm}

\subsubsection{Updating Weights through FL $l$}

We now present the techniques utilized to update the weights within the fully-connected layer. 
The weight update process comprises two steps: firstly, calculating the weight gradients; 
and secondly, removing any accumulated noise in the computed gradient via 
the TEE's re-encryption service. 
However, we aim to minimize the computation workload at TEE, 
as the computational efficiency at TEE 
is comparatively greater than that of REE and
it is also desired to minimize the computation at TEE 
to minimize the chance for side-channel attacks. 
To accomplish this objective, 
we introduce packing mechanisms for the calculated weight gradients. 
These mechanisms reduce the number of ciphertexts that 
need to be processed by TEE for decryption and re-encryption.

Assume the weight gradient is denoted as 
$\tilde{M}_{j, i}^{(l)}$ for $j\in\{0,\cdots,o'_l-1\}$ and $i\in\{0,\cdots,\iota'_l-1\}$, 
where $o'_l$ and $\iota'_l$ represent the number of output ciphertexts and input ciphertexts respectively. 
Then, the weight gradients can be computed by 
the LHE-based multiplication of the output gradients and the encrypted inputs. 
As $n$ inputs are encoded in parallel into one pi-set, 
the gradients for these $n$ inputs need to be aggregated together. 
To accomplish this, 
we calculate the rotation steps using Algorithm~\ref{alg:bwd-cpt-rotation}.
These rotation steps allow us to accumulate the gradients from $n$ inputs into a particular location within $\lceil\log n\rceil$ steps. 
Algorithm~\ref{alg:bwd-weights-full-layer} presents the detailed procedure.

The accumulation of noise from encrypted inputs $\hat{C}_i^{(l)}$ 
and output gradients $\tilde{C}^{(l+1)}_j$
may result in the aggregated weight gradients $\tilde{M}^{(l)}_{j, i}$ at a low LHE level, 
which means further LHE-based multiplication computation could become infeasible. 
To mitigate this issue, we utilize the TEE's re-encryption service for noise reduction (i.e., raising the LHE level of ciphertext). 
To reduce the workload of TEE, we aggregate multiple ciphertexts into one ciphertext by 
using a selector to retrieve the unique summed gradients and 
place them in a unique position within a new ciphertext, 
thereby reducing the total number of ciphertexts fed into TEE. 
The selector is denoted as $\vec U$ in Algorithm~\ref{alg:bwd-weights-TEE-full-layer}, 
which incorporates the learning rate $lr$ and parallel input number $n$ 
to further reduce the required LHE-based multiplication.
The packed weight gradient is denoted as $\tilde{PM}_k^{(l)}$ 
for $k\in\{0,\cdots,\lceil\frac{o'_l\cdot\iota'_l}{n}\rceil-1\}$.
Afterward, TEE can decrypt, decode, re-encode, and re-encrypt the packed weight gradients.
To recover the original data format of the ciphertext, 
we need to spread the packed values in ciphertext, 
which is the reverse process of the previous packing. 
For each weight gradient $\tilde{M}^{(l)}_{j,i}$,
the spreading can be completed in $\lceil\log n\rceil$ steps using Algorithm~\ref{alg:bwd-cpt-rotation}. 
Finally, we can update the weights with the recovered gradients.
The procedure is formally presented in Algorithm~\ref{alg:bwd-weights-TEE-full-layer}.

\subsection{Backpropagation through Convolutional Layer (CL) $l$}

Recall that for the forward propagation in the convolutional layers, 
multiple layers are treated as a combined layer for computation. 
The convolutional operations are performed over the encrypted inputs and the encrypted filters. 
The convolutional output for layer $l$ is provided as the input for layer $l+1$. 
Here, we present our algorithms for the backpropagation over each CL $l$ with encrypted values,
by detailing the step to compute the output gradients and to update the filter weights.

\subsubsection{Computing Output Gradients through CL $l$}

The output gradient for layer $l$ is computed by 
performing convolutional operations on the output gradients in layer $l+1$ and the encrypted filters. 
A detailed procedure for this is provided formally in Algorithm~\ref{alg:bwd-conv-layer}. 
In this algorithm, the output gradient for layer $l$ is denoted as $\tilde{C}^{(l+1)}_{k,u,v}$, 
where $k$ indexes the filters
and $(u, v)$ indexes the 
encrypted inputs in layer $l+1$. 
Additionally, 
each element of an
encrypted filter is denoted as $\hat{F}_{i,x,y}^{(l,k)}$, 
where $i$ is the index of channel,
and $(x,y)$ is the index of 
the element within a filter. 
By applying an LHE-based 
multiplication
to $\tilde{C}^{(l+1)}_{k,u,v}$ and $\hat{F}_{i,x,y}^{(l,k)}$, 
we can derive the output gradient $\tilde{C}^{(l)}_{i,s,t}$ in layer $l$. 
These computed output gradients are then used to calculate the filter gradients.

\begin{algorithm}[htb]
\SetAlgoLined
\caption{Backward Propagation through Convolutional Layer (CL) $l$}
\label{alg:bwd-conv-layer}
\scriptsize
\For{$i\in\{0,\cdots,\alpha_l-1\}$}{ 
    \For{$(u,~v)\in\{0,\cdots,\tilde{\gamma}_{l+1}-1\}^2$}{
        $(u',~v')\leftarrow (\delta_l\cdot u,~\delta_l\cdot v)$\;
        \For{$(x,~y)\in\{0,\cdots,\gamma_l-1\}^2$}{
            \For{$k\in\{0,\cdots,\epsilon_l-1\}$}{ 
                \If{not visited $(u'+x, v'+y)$}{
                $\tilde{C}^{(l)}_{i,u'+x,v'+y}\leftarrow 0$;
                \Comment{initialize each output}\;
                }
                $\tilde{C}^{(l)}_{i,u'+x,v'+y}\oplus = \tilde{C}^{(l+1)}_{k,u,v}\otimes\hat{F}^{(l,k)}_{i,x,y}$;
            }
        }
        
    }
}

\end{algorithm}

\begin{algorithm}[htb]
\SetAlgoLined
\caption{Updating Kernel Gradients through Convolutional Layer (CL) $l$}
\label{alg:bwd-kernals-conv-layer}
\scriptsize
\For{$k\in\{0,\cdots,\epsilon_l-1\}$}{ 
    \For{$i\in\{0,\cdots,\alpha_l-1\}$}{
        \For{$(x,~y)\in\{0,\cdots,\gamma_l-1\}^2$}{
            $\tilde{F}^{(l,k)}_{i,x,y}\leftarrow 0$;
            \Comment{initialize each output}\;
            \For{$(u,~v)\in\{0,\cdots,\tilde{\gamma}_{l+1}-1\}^2$}{
                $(u',~v')\leftarrow (\delta_l\cdot u,~\delta_l\cdot v)$\;
                $\tilde{F}^{(l,k)}_{i,x,y}\oplus = \hat{C}^{(l)}_{i,u'+x,v'+y}\otimes\tilde{C}^{(l+1)}_{k,u,v}$;
            }
            $p \leftarrow (k\cdot\alpha\gamma_l^2+i\cdot\gamma_l^2+x\cdot\gamma_l+y)\%n$\;
            $\vec{R} \leftarrow computeRotations(p, n)$; 
            \Comment{compute the rotate directions and steps to sum up gradients}\;
            \For{$t\in\{0,\cdots,\lceil\log n\rceil\}$}{
            \Comment{sum up gradients from $n$ parallel images}\;
                $\tilde{F}^{(l,k)}_{i,x,y} \oplus= Rot(\tilde{F}^{(l,k)}_{i,x,y}, 2^t\cdot\vec{R}_t)$;
            }
        }
    }
}

\end{algorithm}

\begin{algorithm}[htb]
\SetAlgoLined
\caption{Removing Accumulated Noise of Kernels in Convolutional Layer (CL) $l$}
\label{alg:bwd-kernals-TEE-conv-layer}
\scriptsize
\For{$k\in\{0,\cdots,\epsilon_l-1\}$}{ 
    \For{$i\in\{0,\cdots,\alpha_l-1\}$}{
        \For{$(x,~y)\in\{0,\cdots,\gamma_l-1\}^2$}{
            \Comment{pack kernal gradients $\tilde{F}^{(l,k)}_{i,x,y}$ together}\;
            $idx \leftarrow k\cdot\alpha\gamma_l^2+i\cdot\gamma_l^2+x\cdot\gamma_l+y$\;
            $(p, t) \leftarrow (idx\%n, idx/n)$\;
            $\vec{U} \leftarrow [0,\cdots,\beta,\cdots,0,\cdots,\beta,\cdots]$;
            \Comment{for position $\{p,p+n,p+2n,\cdots,p+\lfloor\frac{{\cal S} -p}{n}\rfloor\cdot n\}$, $\vec{U}$ has value $\beta=\frac{lr}{n}$ where $lr$ is learning rate, otherwise 0}\;
            \If{not visited $t$}{
                $\tilde{PF}^{(l)}_t\leftarrow 0$;
                \Comment{initialize each output}\;
            }
            $\tilde{PF}^{(l)}_t\oplus=\tilde{F}^{(l,k)}_{i,x,y}\otimes\vec{U}$\;
        }
    }
}
\For{$k\in\{0,\cdots,\lceil\frac{\epsilon_l\cdot\alpha_l\cdot\gamma_l^2}{n}\rceil-1\}$}{
    \Comment{decrypt, decode, re-encode and re-encrypt the packed weights in TEE}\;
    $\vec{pt}\leftarrow Dec_{sk}(\tilde{PF}^{(l)}_k)$;
    $\tilde{PF}^{(l)}_k \leftarrow Enc_{pk}(\vec{pt})$;
}
\For{$k\in\{0,\cdots,\epsilon_l-1\}$}{ 
    \For{$i\in\{0,\cdots,\alpha_l-1\}$}{
        \For{$(x,~y)\in\{0,\cdots,\gamma_l-1\}^2$}{
            \Comment{unpack weights $\tilde{PF}^{(l)}_t$ and recover $\tilde{F}^{(l,k)}_{i,x,y}$}\;
            $idx \leftarrow k\cdot\alpha\gamma_l^2+i\cdot\gamma_l^2+x\cdot\gamma_l+y$\;
            $(p, t) \leftarrow (idx\%n, idx/n)$\;
            $\vec{U} \leftarrow [0,\cdots,1,\cdots,0,\cdots,1,\cdots]$\;
            \Comment{for position $\{p,p+n,p+2n,\cdots,p+\lfloor\frac{{\cal S} -p}{n}\rfloor\cdot n\}$, $\vec{U}$ has value 1, otherwise 0}\;
            $\tilde{F}^{(l,k)}_{i,x,y}\leftarrow\tilde{PF}^{(l)}_t\otimes\vec{U}$\;
            $\vec{R} \leftarrow computeRotations(p, n)$; 
            \Comment{compute the rotate directions and steps to spread gradients}\;
            \For{$s\in\{0,\cdots,\lceil\log n\rceil\}$}{
                $\tilde{F}^{(l,k)}_{i,x,y} \oplus= Rot(\tilde{F}^{(l,k)}_{i,x,y}, -2^s\cdot\vec{R}_s)$;
            }
            $\hat{F}^{(l,k)}_{i,x,y} \oplus= \tilde{F}^{(l,k)}_{i,x,y}$;
            \Comment{update kernal with gradients}
        }
    }
}
\end{algorithm}

\subsubsection{Updating Filters through CL $l$}

To compute the filter gradients in layer $l$, 
we perform convolutional operations over the output gradients in layer $l+1$ and the inputs in layer $l$. 
However, 
the convolutional operation increases noises quickly and 
may prevent the resulting ciphertexts from performing more multiplication operations. 
To address this, 
the refining service at the REE can interactively requests the TEE's re-encryption service to reduce the noises
while minimizing TEE's involvement.

To aggregate the gradients $\tilde{F}^{(l,k)}_{i,x,y}$ from $n$ parallel inputs and 
add them to a specific location determined by the index $(k, i, x, y)$, 
we need to perform the operation in $\lceil\log n\rceil$ steps, 
which can be achieved using Algorithm~\ref{alg:bwd-cpt-rotation}. 
The details of computing and aggregating the gradients are presented formally in Algorithm~\ref{alg:bwd-kernals-conv-layer}.

To reduce the 
computation
workload of TEE, 
the filter gradients are aggregated at different slots of a vector and then packed together before being sent to TEE for noise removal. 
This packing step allows us to further aggregate multiple ciphertexts into one ciphertext, 
by multiplying each ciphertext with a selector to retrieve the valid aggregated values 
and store them at different slots of the new ciphertext. 
The formal description of this step can be found in Algorithm~\ref{alg:bwd-kernals-TEE-conv-layer} from line 1 to line 14. 
After the packed ciphertexts are sent to TEE, 
they undergo a process of decryption, decoding, re-encoding, and re-encryption before being sent back to REE for unpacking and recovering. 
The unpacking and recovering steps are the reverse of the packing process. 
Once the recovering step finishes, 
the filter weight can be updated with the gradients.

\section{Evaluation}
\label{sec:eval}

In this section, 
we present the experimental setup for
the prototype of
our proposed autonomous and confidential model refining scheme.
We demonstrate the feasibility and effectiveness of our scheme through 
evaluation results using various metrics, 
such as computation time, 
communication cost between TEE and REE, and 
training accuracy compared to plaintext training.

\subsection{CNN Models and LHE Parameters}

In our experiments, we utilized a CNN model with $c=2$ and $f=2$, 
while training on the MNIST~\cite{lecun1998mnist} dataset 
to classify handwritten digits in images of dimensions $28\times 28$. 
Our CNN model trains batches of $n=128$ inputs at a time. 
For the convolutional layer (CL) $c=1$ and $c=2$, 
we use $(4, 3, 3)$ and $(4, 2, 1)$ as 
filter numbers, filter side, and filter stride. 
We set the number of output neurons to 32 and 10 for FL $f=1$ and $f=2$, respectively. 
Initially, we train a base model on data dominated by odd-labeled images, 
i.e., those with labels $\{1,3,5,7,9\}$, 
along with a small number of even-labeled images, 
i.e., those with labels $\{0,2,4,6,8\}$, 
at a ratio of odd to even images of $100:1$. 
Following this, we refine the base model with a smaller dataset of a different distribution, 
where the even-labeled images are dominant. 
We compare the performance of the refined model trained using 
our proposed confidential training scheme 
with that of training in 
plaintexts directly
for a few epochs, 
to demonstrate the feasibility of our system and the accuracy-changing tendencies of our approach.

The LHE scheme we utilize is CKKS~\cite{cheon2017homomorphic} implemented in SEAL~\cite{sealcrypto} library. 
To enable one forward and backward propagation process, 
we select a polynomial modulus of $N=16384$ with a maximal coefficient modulus bit length of 438, 
ensuring a 128-bit security level. 
Each plaintext/ciphertext can encode up to ${\cal S} = \frac{N}{2} $ values, 
where $\cal S$ represents the slot number. 
The detailed modulus parameters are $\{40, 30, 30, 30, 30, 30, 30, 30, 30, 30, 30, 40\}$, 
which can support up to ${\cal L}=10$ levels of consecutive LHE-based multiplication. 
In each forward propagation, each CL, fully-connected layer (FL), and activation layer 
requires one level of homomorphic multiplication. 
In backpropagation, we resume the output gradients for the last fully connected layer $f-1$ to the highest level ${\cal L}-1$, 
as TEE is 
employed for re-encryption
to remove the accumulated noise.

\subsection{Evaluation Results}
The evaluation is performed with a laptop, i.e., a Macbook Air equipped with an Intel 1.6 GHz CPU and 8 GB of memory. 
Note that, the cloud server in the proposed scheme is run in a single thread employing only one CPU core,
and no GPU is utilized. 

\subsubsection{Computation Time for Each Batch}

We evaluate the computation time for each layer in the forward propagation and backpropagation process, respectively,
where each convolutional layer is followed by a square activation function. 
Table~\ref{tab:forward} shows 
the number of homomorphic operations and the incurred execution time for the forward propagation. 
Note that, we have demonstrated only the homomorphic multiplication and rotation operations 
since the execution time for homomorphic addition and multiplication with plaintext is comparatively smaller. 
For the CLs, the combined stride is computed as 
$\tilde{\gamma}_0=6$ and $\tilde{\gamma}_1=2$ based on the filter side and stride. 
The number of multiplications for each CL layer $l$ is 
$\tilde{\gamma}_l^2\times \epsilon_l\times \alpha_l$, 
which is 144 and 64, as indicated in table~\ref{tab:forward}. 
In CL2, 64 neurons are packed into one ciphertext, and 
these ciphertexts, derived from different filters, are fed as inputs of FL1 (Type I inputs). 
The major execution time for FL1 is from multiplication and rotation. 
The number of input ciphertexts is 4, and the number of output ciphertexts is 32, leading to 128 multiplications. 
The rotation is computed in $\lceil\log 64 \rceil\times32=192$ steps. 
The execution time for the forward propagation is mainly due to the computation of CL1 and FL1. 
In CL1, homomorphic addition and multiplication with plaintext are also involved. 
The total time for each forward propagation of the refining process is 34.12 seconds. 

\begin{table}[htb]
\centering
\caption{Execution Time (unit: second) of One Forward Propagation Round}
\label{tab:forward}
\begin{tabular}{|c|cc|c|}
\hline
       & \multicolumn{2}{c|}{Count of LHE Operations}  & \multirow{2}{*}{Incurred Computational Time} \\ \cline{1-3}
Layer  & \multicolumn{1}{c|}{$\otimes$} & Rot &                       \\ \hline
CL1    & \multicolumn{1}{c|}{144}       & -   & 13.011                \\ \hline
CL2    & \multicolumn{1}{c|}{64}        & -   & 3.859                 \\ \hline
FL1    & \multicolumn{1}{c|}{128}       & 192 & 14.161                \\ \hline
FL2    & \multicolumn{1}{c|}{32}        & -   & 3.089                 \\ \hline
\end{tabular}
\end{table}

\begin{table*}[htb]
\centering
\caption{Execution Time (unit: second) for One Backpropagation Round}
\label{tab:backward}
\begin{tabular}{|c|c|cccc|c|}
\hline
\multirow{2}{*}{} & \multirow{2}{*}{Output Gradients} & \multicolumn{4}{c|}{Weight Update}                                                                                   & \multirow{2}{*}{Execution Time per Layer} \\ \cline{3-6}
                  &                                   & \multicolumn{1}{c|}{Weight Gradients} & \multicolumn{1}{c|}{Packing} & \multicolumn{1}{c|}{Decrypt/Encrypt} & Spreading &                                           \\ \hline
FL2               & 0.104                             & \multicolumn{1}{c|}{6.3}              & \multicolumn{1}{c|}{0.886}   & \multicolumn{1}{c|}{0.186}           & 19.923 & 27.399                                    \\ \hline
FL1               & 15.235                            & \multicolumn{1}{c|}{28.033}           & \multicolumn{1}{c|}{3.527}   & \multicolumn{1}{c|}{0.092}           & 46.522 & 93.409                                    \\ \hline
CL2               & 12.235                            & \multicolumn{1}{c|}{16.975}           & \multicolumn{1}{c|}{2.494}   & \multicolumn{1}{c|}{0.091}           & 24.528 & 56.323                                    \\ \hline
CL1               & 6.564                             & \multicolumn{1}{c|}{6.155}            & \multicolumn{1}{c|}{0.94}    & \multicolumn{1}{c|}{0.082}           & 14.597 & 28.338                                    \\ \hline
\end{tabular}
\end{table*}

Table~\ref{tab:backward} presents an analysis of the execution time incurred in the backpropagation of a single batch. 
Once the forward propagation pass is completed, 
the output is provided to the TEE to decrypt, compute the softmax function with loss and output gradients, and re-encrypt. 
These computations are carried out over plain values in TEE and thus is fast. 
The weight update process comprises four steps, namely, 
computing weight gradients, packing gradients, emplying TEE for re-encryption, and spreading weights. 
According to the findings in Table~\ref{tab:backward}, 
spreading the weights in each layer is the most time-extensive step. 
This is because the $\emph{spread}$ step involves the unpacking process, 
which entails multiplying with the selector, 
the reverse step of packing, 
rotating the ciphertext to recover values in all slots, and 
updating the original weight with spread gradients. 
Additionally, the computation necessary for spreading is performed over gradients at a higher encryption level, 
leading to higher computation time than the weight gradients or output gradients. 
Furthermore, the decryption and encryption steps for TEE take significantly less time than other steps in REE, 
as demonstrated in Table~\ref{tab:backward}.
Overall, the total execution time for the backward propagation of refining process is 205.47 seconds.

\begin{table}[htb]
\centering
\caption{Communication Cost between Clients and the Cloud Server}
\label{tab:communication}
\resizebox{0.8\linewidth}{!}{
\begin{tabular}{|c|c|c|}
\hline
               & Plaintext Model & Ciphertext Model \\ \hline
Model Provider & 176KB           & 435MB            \\ \hline
Data Provider  & 924KB           & 543MB            \\ \hline
\end{tabular}
}
\end{table}

\subsubsection{Communication Cost between Clients and REE} 

We measure communication cost between the client, which can be either the model provider or data provider, 
and the cloud server, 
for both our proposed confidential refining scheme and the baseline scheme that works on plaintext data/model.
The model provider sends the model to the server, 
while the data provider shares the inputs with the server either in plaintext or encrypted format. 
For our proposed confidential scheme, 
we use 1152 images for training, and the model is a CNN with 2 CLs and 2 FLs. 
The results in Table~\ref{tab:communication} show that 
encrypted data incurs a higher communication overhead than plain data. 
However, it is important to note that the communication between the clients and the server occurs only at the initial stage. 
Once the encrypted model and data are received by the server, 
no further communication is required during the refining process.

\begin{table}[htb]
\centering
\caption{Test accuracy of Model Refining. (Note that, the Baseline Scheme refines plaintext model with plaintext data 
while Our Scheme refines encrypted model with encrypted data.)}
\label{tab:accuracy}
\resizebox{\linewidth}{!}{
\begin{tabular}{|c|c|c|c|c|c|}
\hline
                 & Epoch 1 & Epoch 2 & Epoch 3 & Epoch 4 & Epoch 5 \\ \hline
Baseline Scheme  & 0.7014  & 0.7839  & 0.8255  & 0.8438  & 0.8559  \\ \hline
Our scheme & 0.4983  & 0.6858  & 0.7387  & 0.7465  & 0.7517  \\ \hline
\end{tabular}
}
\end{table}

\subsubsection{Accuracy of Model Refining}

To demonstrate the feasibility of our scheme, 
we compare the accuracy of the models refined by our proposed scheme and by the baseline scheme over plaintext data/model.

We first prepare a base model.
It is trained based on 10000 odd-labeled and 100 even-labeled images.
Ten epochs of training has been conducted to the get the base model with a training accuracy of 96.51\%.
Then we prepare a new batch of data for refining,
which consists of 1152 images, 
including 1000 images with even labels and 152 images with odd labels.
As we can see, the two data sets have different distributions and thus
the base model should not fit well with the refining data set. 
In fact, with a test data set having same distribution as the new data set for refining,
the base mode has a testing accuracy of 48.09\%.
On top of the base model,
we employ both our proposed confidential refining scheme and 
the baseline scheme that refines plaintext-model with plaintext-data. 
The results of these two schemes are shown in Tabel~\ref{tab:accuracy}. 
As we can, the testing accuracy for these two scheme both increase as the training epoch,
and the improvements in accuracy are comparable. 
The baseline scheme achieves higher accuracy for each epoch, 
due to the optimization methods used in PyTorch's default training process 
that are hard to be completely implemented in our confidential scheme. 
However, the results demonstrate the feasibility of autonomous and continuous refining. 


\section{Related Works}



Extensive research has been focused on the confidentiality-preserving deep neural network inference. 
Generally, these research works are mainly based on the following techniques: 
homomorphic encryption, multi-party computation, trusted execution environment, 
or a combination of these techniques.



Schemes that utilize leveled homomorphic encryption 
includes~\cite{gilad2016cryptonets,jiang2018secure,bourse2018fast,chou2018faster,dathathri2019chet,xie2021privacy}.
As one of the early efforts,
the CryptoNets proposed by Gilad-Bachrach et al.~\cite{gilad2016cryptonets} 
employs the packing technique~\cite{simd}
to efficiently conduct inference using encrypted data over a plaintext CNN model. 
Among them, CryptoNets~\cite{gilad2016cryptonets} is one of the first works 
applying the packing technique~\cite{simd} for inference based on a CNN model.
As the proposed technique packs one value from each input into a ciphertext,
a large number of inputs can be processed in parallel and thus 
a high level of amortized efficiency can be attained. 
Jiang et al.~\cite{jiang2018secure} propose E2DM,
which packs a matrix into a ciphertext and bases on this to efficiently multiply two encrypted matrices. 
Such techniques have been applied to inference using encrypted data
over an encrypted CNN model with only one convolutional layer. 
Later, Xie et al.~\cite{xie2021privacy} propose PROUD,
which combines the packing techniques and parallel execution to further speed up the inference
based on encrypted data and model. 
%
Recently, Liu and Zhang~\cite{liu2022towards} further designs a packing-based inference scheme
that can be applied for more generic CNN models and is shown to be flexible to the number of available parallel inputs. 

SMPC-based schemes have been extensively studied recently 
as they are more computationally-efficient than the schemes based on homomorphic encryption.
Among them, most~\cite{mohassel2017secureml,rouhani2018deepsecure,liu2017oblivious,juvekar2018gazelle,mishra2020delphi,chandran2019ezpc,xie2019bayhenn,riazi2019xonn,riazi2018chameleon,lu2017crypti,xie2021privacy, huang2022cheetah} 
are based on two-party computation, with which
the client and server need to interact with each other 
while the computation is being performed. 
For example, 
in the NiniONN scheme proposed by Liu et al.~\cite{liu2017oblivious}
and the GAZELLE scheme proposed by Juvekar et al.~\cite{juvekar2018gazelle},
the inputs are split between the client and server as additive secret shares and
non-linear computation is implemented with garbled circuits for confidentiality. 
The GAZELLE is extended by Mishra et al.~\cite{mishra2020delphi} to design the DELPHI scheme,
which is hybrid scheme that generates the neural network architecture configurations
to strike the balance between performance and accuracy. 
Researchers have also proposed schemes based on three-party computation
\cite{mohassel2018aby3, wagh2020falcon, patra2020blaze, koti2021swift} or four-party computation
\cite{chaudhari2019trident, byali2019flash, koti2021swift},
which however assume that a majority of the parties are honest. 
Overall, the requirement of frequent interactions among the multiple parties in these schemes
could cause high communication overheads and latency.

There have also been schemes~\cite{tramer2018slalom, zhang2021citadel, natarajan2021chex}
proposed by leveraging trusted execution environment technologies such as Intel SGX. 
For example, Zhang et al.~\cite{zhang2021citadel} propose a system named Citadel.
In this system, code is partitioned to two parts:
data handling code executed by multiple training enclaves and 
model handling code executed by an aggregation enclave.
%
%
Natarajan et al.~\cite{natarajan2021chex} propose the CHEX-MIX system,
which combines the homomorphic encryption and TEE for
the confidentiality of data and model and for the integrity of computation.
%
%
As discussed in Section I,
the TEE based schemes may not fully utilize the memory and computation resources 
and may suffer from side-channel attacks. 



Compared to the related works, 
our proposed solution is unique in the following aspects:
    In terms of the application settings, our proposed solution is to protect the confidentiality of
    data and model for autonomous and continuous model refining in cloud, while
    the related works were developed to protect the confidentiality of data or model
    mostly for the model-based inference and some for model training. 
    In terms of techniques adopted, our proposed solution leverages 
    both leveled homomorphic encryption and TEE in a complementary manner.
    Here, most of the computation is conducted over homomorphically-encrypted data and model parameters;
    the TEE is used only for key management, and 
    for re-encryption that periodically reduce the noises from the ciphertexts 
    to make the model refining process sustainable. 

\section{Conclusions and Future Work}
\label{sec:conclusion}


In this paper, 
we design and implement a scheme that 
enables autonomous and confidential model refining in cloud,
based on the integration of leveled homomorphic encryption
and trusted execution environment technology. 
Specifically, 
the cloud server has a trusted execution environment (i.e., TEE) that
provides the initialization service and 
the homomorphically re-encryption (i.e., noise reduction) service, and
a model refining service running in the regular (untrusted) execution environment (i.e., REE).
A client can join the system by providing a base model homomorphically-encrypted
with a key obtained from the TEE that it has attested. 
The same or a different client can further provide a new batch of training data
homomorphically-encrypted with a key obtained from the TEE as well, 
every now and then. Upon receiving the new encrypted training data,
with assistance from the TEE, 
the refining service can refine the current model based on the new data
autonomously without accessing to the data/model in plaintext or 
intervention from the clients. 
Experiments have been conducted to demonstrate the feasibility of the scheme. 
However, the computational efficiency of the scheme is still significantly lower than 
the baseline scheme that refines plaintext-model with plaintext-data.
In the future, we plan to improve the performance of the scheme 
by utilizing higher level of parallelism and GPU at the cloud server.

\bibliographystyle{IEEEtran}
\scriptsize\bibliography{sample}

\end{document}